# Some Applications of Isotope - Based Technologies: Human Health and Quantum Information.

Vladimir G. Plekhanov

Computer Science College, Erika 7a, Tallinn, 10416, ESTONIA.

**Abstract.** *Technology is the sum of the information, knowledge and agency. This takes energy and information as fundamental concepts. In this paper I'll try to describe very briefly in popular form of some applications of radioactive and stable isotopes in medicine and quantum information, respectively.*

Research into the use of isotopes for medical, industrial, environmental, and other important science applications shows great promise to improve the quality of life for the citizens of separated countries and throughout the world. Currently, more than 12 million nuclear medicine procedures are performed each year in the United States, and it is estimated that one in every three hospitalized patients has a nuclear medicine procedure performed in the management of his or her illness.

Isotopic tracer methods find applications in nearly every field of science, e.g. medicine, biology, physiology, nutrition, toxicology, biotechnology, which are typically life science fields, or more technical areas, as physics, chemistry, agriculture, geoscience, engineering, which have now become integral part of every day life (see, for example [1, 2]). There are many isotopes that are used in medicine. For example, $^{60}$Co is used to produce beam radiation that is used in killing cancer cells. At first, the radio - isotopes utilized were naturally occurring ones such as radium - 226, radium - 224, radon - 222, polonium - 210, tritium (hydrogen - 3), carbon - 14, etc. Even today, "radium needless" and "radon seeds" are used to shrink cancerous tumors. Along with cancer therapy (oncology), a number of isotopes are used, as noted above, for medical imaging. These radioactive isotopes are injected into the body, and then imaging devices follow the path of these isotopes through the body. Some of the tracer isotopes indicate below in the Table 1 (for details see, also [3]). The images below were taken using positron emission tomography (PET) (see e.g. [4-6]). This technique scans for positrons emitted by several isotope, including $^{11}$C and $^{10}$F (see, also Table 1). This particular image shows brain activity of a patient with Parkinson's disease (Fig. 1) (for details see [5]).

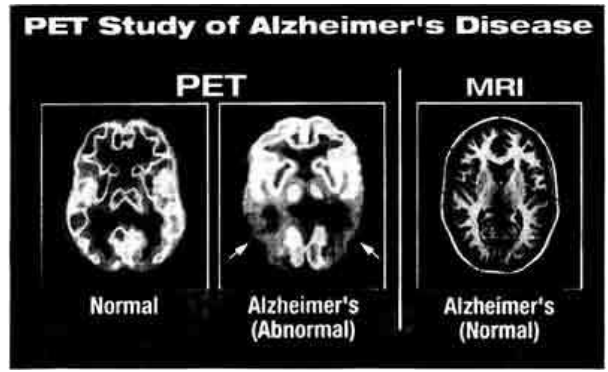

**Fig. 1. Present picture (PET) shows brain activity of a patient with Parkinson's disease (see, also, text).**

This part of my paper is not intended to cover all developments in the quantum information theory and quantum computation. My aim is rather to provide the necessary insights for an understanding of the field so that various nonexperts can judge its fundamental and practical importance. Quantum information theory and quantum communication and computation are an extremely exciting and rapidly growing field of investigation (see, e.g., [7 - 10] and references therein). As is well-known, information is not a disembodied abstract entity: it is always tied to the physical representation (see, e.g. [11]). It is represented by engraving on a stone tablet, a spin, a charge, a hole, in a punched card, a mark on paper, or some other equivalent. This ties the handling of information to all the possibilities and restrictions of our real physical word, its laws of physics and, its storehouse of available parts. Indeed, our assertion that information is physical amounts to an assertion that mathematics and computer science are a part of physics (see, also [12]). Mathematicians, in particular, have long assumed that mathematics was there first, and that physics needed that to describe the Universe. The achievements, over the last four decades or so, in atomic and laser physics, as well as measured techniques of experimental physics [13] have developed in the application of quantum theory to individual systems (electrons, atoms, ions, etc.) and to mesoscopic or even macroscopic systems where a small number of collective degrees of freedom show genuine quantum behavior. One exciting aspect of this developing fundamental research is its technological potential. It could span what might be termed quantum information technology. In such a scenario, machines would process and exchange information according to the laws of quantum physics, in contrast to the workings of conventional information technology, where all this is done classically. Information processing now plays a significant role in all of our lives. We carry and use an increasing number of cards containing magnetically stored data. Many household and workplace appliances contain processing power, from simple microprocessors through to powerful computers. It is obvious that quantum physics is not going to make significant inroads into this huge technology spectrum in the foreseeable future. Nevertheless, even if quantum machines could outperform their classical counterparts (or, better still, open up completely new avenues) in just a few useful applications, there would be real excitement. Quantum engineering would begin to evolve (see, also [14]).

Apart from the computational power of a quantum computer there is a much more banal argument for incorporating quantum mechanics into computer science: *Moore's* *law*. In 1965 Intel co-founder Gordon Moore observed an exponential growth in the number of transistors per square inch on integrated circuit and he predicted that this trend would continue [15]. In fact [16[a]], since then this density has doubled approximately every 18 months (see, also Fig. 7 in [16[b]]). If this trend continuous then around the year 2020 the components of computers (gates) are at the atomic scale where, naturally, quantum effects are dominant (for details, see [9]). Let us now have look at the way a quantum computer works. A classical computer its input according to its program to produce the output. Any classical system is always on one a defined set of states. For example, a perfect classical bit is actually in state zero or state one at any time; the two possibilities are mutually exclusive (see, also [11] and references therein). However, a quantum system can exist in what might be termed a schizophrenic state, known as a superposition state [18]. Quantum computation and quantum information are built upon analogous concept, the quantum bit, or qubit for short [8]. It is a two-dimensional quantum system (for example, a spin 1/2, a photon polarization, an atomic system two relevant states, etc.) with Hilbert space. In mathematical terms, the state of quantum state (which is usually denoted by $|\Psi>$ [17]) is a vector in an abstract Hilbert space of possible states for the system.

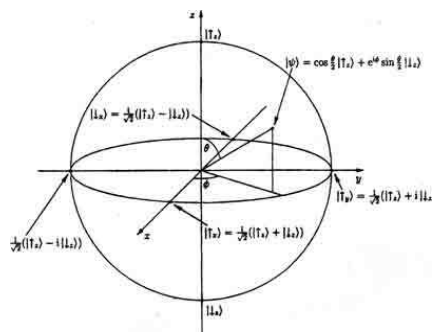

**Fig. 2. Bloch sphere representation of a qubit (after [8]).**

The space for a single qubit is spanned by a basis consisting of the two possible classical states, denoted, as above, by $|0>$ and $|1>$. This mean that any state of qubit can be decomposed into the superposition

$$|\Psi> = \alpha |0> + \beta |1> \quad (1)$$

with suitable choices of the complex coefficients *a* and *b*. The value of a qubit in state $|\Psi>$ is uncertain; if we measure such a qubit, we cannot be sure in advance what result we will get. Quantum mechanics just gives the probabilities, from the overlaps between $|\Psi>$ and the possible outcomes. Thus the probability of getting 0 is $|<0|\Psi>|^2 = |a|^2$ and that for 1 is $|<1|\Psi>|^2 = |b|^2$. Quantum states are therefore normalized; $<\Psi|\Psi> = (b^* a^*) \cdot \begin{pmatrix} b \\ a \end{pmatrix} = 1$ (where $|\Psi>$ is represented by the vector $\begin{pmatrix} b \\ a \end{pmatrix}$) and the probabilities sum to unity. Quantum mechanics also tells us that (assuming the system is not absorbed or totally destroyed by the action of measurement) the qubit state of Eq. (1) suffers a projection to $|0> (|1>)$ when we get the result 0(1). Because $|\alpha|^2 + |\beta|^2 = 1$ we may rewrite Eq. (1) as (see, e.g. [8])

$$|\Psi> = \cos\theta\,|\,0> + e^{i\varphi}\sin\theta\,|\,1> \qquad (2)$$

where $\theta$, $\varphi$ are real numbers. Thus we can apparently encode an arbitrary large amount of classical information into the state of just one qubit (by coding the information into the sequence of digits of $\theta$ and $\varphi$). however in contrast to classical physics, quantum measurement theory places severe limitations on the amount of information we can obtain about the identity of a given quantum state by performing any conceivable measurement on it. Thus most of the quantum information is "inaccessible" but it is still useful - for example it is necessary in its totality to correctly predict any future evolution of the state and to carry out the process of quantum computation (see, e.g. [8]).

The numbers $\theta$ and $\varphi$ define a point on the unit three - dimensional sphere, as shown in Fig. 2. This sphere is often called the Bloch (Poinkare) sphere [8]; it provides a useful means of visualizing the state of a single qubit. A classical bit can only sit at the north or the south pole, whereas a qubit is allowed to reside at any point on the surface of the sphere (for details see, also [9]).

Besides the quantum computer with its mentioned applications quantum information science yields a couple of other useful applications which might be easier to realize. The best example is quantum cryptography (see, e.g. [18]) which enables one to transmit information with the security of nature's laws [9]. But for the first I should give the definitions of some technical words:

1. Cryptology: The study of secure communications, which involves both cryptography and cryptoanalysis.

2. Cryptography: It's an art and science of using mathematics to secure information and create a high degree of trust in the electronic realm.

3. Cryptoanalysis: The branch of cryptology dealing with breaking of cipher to recover information, or forging encrypted information that will be accepted as authenic.

The word "cryptography" is derived from the Greek 'kryptos' (hidden) and 'graphia' (writing). The dictionary, defines cryptography as hidden writing.

The answers the question, "Who needs encryption?", one can examine the various reasons why banks, businesses, professionals, military, everyday people, and even criminals require some sort of protection. Cryptography is used whenever someone wants to send a secret message to someone else, in situation where anyone might be able to get hold of the message and read it. Cryptography provides a solution to the problem of information security and privacy. For electronic communications, the techniques of private and public key cryptography are becoming increasing popular. Cryptography provides integrity i.e. assures that the information was not modified while in transit. Identification and authentication are two widely used applications of cryptography. Identification is the process of verifying someone's or something's identity. Authentication merely determines whether that person or entity is authorized for whatever is in question. For this purpose digital Signatures are used.

Quantum cryptography is on leading edge of cryptographic implementations. It is currently relegated to the laboratory for reasons of technical feasibility. Signal, for example [18], have a certain polarization, as long as the polarization remains unchanged, the signal has not been intercepted or monitored in any way. Interception or monitoring causes a polarization shift. Quantum cryptography uses this technology to publicly distribute key information (see, for example [19]). The receiver records a polarization and asks the sender the sender if the recorder polarization is correct. If it is, then the receiver knows it has a valid key unknown to anyone else. The Heisenberg principle states [18] that a state cannot be monitored without changing the state itself. So far, on quantum level, this is true. This means that if the key is monitored during transmission, the polarization will change, and the sender will detect this because the polarization information returned from receiver will be in-correct.

I should add that cryptography is not confined to the world of computers. Cryptography is also used in mobile phones as a means of authentication; that is, it can be used to verify that a particular phone has the right to bill to a particular phone number. This prevents people from cloning mobile phone numbers and access codes. Another application is to protect phone calls from eavesdropping using voice encryption.

Returning to quantum computer, it is necessary to note that the choice of architecture of processor will be crucial to a first demonstration of solid state quantum computer. Kane original proposal [20] envisions encoding quantum information onto the nuclear spin 1/2 states of $^{31}$P qubits in a spinless I = 0 $^{28}$Si lattice. Nuclear spin relaxation times for $^{31}$P donors are extremely long when the electron spin is polarized, many hours at LHeT temperature and far longer below. The Kane architecture employs an array of top-gates to manipulate the ground state wavefunctions of the spin-polarized electrons at each donor site in a high magnetic field B ~ 2T, at very low temperature T ~ 100 mK. Electrons spins on isolated Si;P donors have very long decoherence times of ~ 60 ms in isotopically purified $^{28}$Si at 7 K [21]. To concluding I would like to point out once more possibility to architecture of quantum computer: a new NMR quantum computer made exclusively of mixed crystals of LiH$_x$D$_{1-x}$ [22] with possibility of a strong nuclear polarization enhancement leading to much improved scalability [23].

**Table. Some very often used in everyday life radioactive isotopes (see also text)**

| Isotope | Half-live | Applications |
|---|---|---|
| Actinium -225 | 10.0 d | An alpha emitter that shows promisw in the treatment of certain types of cancer |
| Californium - 252 | 2.64 y | Used to treat cervical cancer melanoma, brain cancer treatment |
| Cobalt - 60 | 5.27 y | A gamma emetter used in irradiations food and medical equipment sterilization |
| Tungsten - 188 | 69.8 d | Used to prevent the re-closure (restenosis) of coronary arteries following heart surgery |
| Copper -67 | 61.9 h | Used to label monoclonal antibodies and destroy tatget tumors; PET scanning |
| Strontium - 82 | 25.6 d | PET scanning |
| Calcium - 42 | Stable | Along with calcium - 44 used in human calcium retention studies |
| Lithium - 6 | Stable | Neutron capture terapy research |
| Carbon - 11 | 20.3 m | Radiotracer in PET scans to study normal/abnormal brain function |
| Germanium - 68 | 271 d | PET imaging |
| H - 3 | 12.3 y | Labeling PET imaging |